\documentclass[preprint,12pt]{revtex4}
\usepackage{graphicx}% Include figure files

\begin{document}

\title[Kondo physics and superconductivity]{Kondo resonance enhanced supercurrent in single wall carbon nanotube Josephson junctions}

\author{K Grove-Rasmussen}
\email{k_grove@fys.ku.dk}
\author{H Ingerslev J\o rgensen and P E Lindelof}

\affiliation{Nano-Science Center, Niels Bohr Institute, University
of Copenhagen, Denmark}

\begin{abstract}
We have contacted single wall carbon nanotubes grown by chemical
vapor deposition to superconducting Ti/Al/Ti electrodes. The device,
we here report on is in the Kondo regime exhibiting a four-fold
shell structure, where a clear signature of the superconducting
electrodes is observed below the critical temperature. Multiple
Andreev reflections are revealed by sub-gap structure and a narrow
peak in the differential conductance around zero bias is seen
depending on the shell filling. We interpret the peak as a proximity
induced supercurrent and examine its interplay with Kondo
resonances.
\end{abstract}

\maketitle

\section{Introduction}
Single wall carbon nanotubes (SWCNT) have been under intense
investigation for more than a decade due to their unique mechanical
and electrical properties. They are one-dimensional conductors with
two conducting modes and when contacted to electrodes they behave as
quantum dots, where phenomena as Fabry-Perot interference
\cite{Liang}, Kondo effect \cite{Nygaard} and Coulomb blockade
\cite{tans, bockrath} have been observed. The different regimes can
be accessed by indirectly controlling the coupling between the SWCNT
and the contacts by choice of contact material. In the low
transparency regime (closed quantum dot), electron transport is
blocked except at charge degeneracy points, where electrons can
tunnel through the SWCNT only one by one due to Coulomb blockade.
When the transparency is increased (intermediate regime) cotunneling
of electrons becomes possible, which can give rise to the Kondo
effect. Finally in the high transparency regime electrons on the
SWCNT are not well defined and the phenomena observed (broad
resonances) are due to interference of electron waves (Fabry-Perot
interference). The possibility of contacting carbon nanotubes to
superconducting leads \cite{Kasumov,Morpurgo,Krstic2003,haruyama}
opens up for the interesting study of effects related to
superconductivity such as supercurrent and multiple Andreev
reflections \cite{Buitelaar} together with the above mentioned
phenomena. In the Fabry-Perot regime recent experiments have
confirmed that the supercurrent is modulated by the quantized nature
of the energy spectrum of the SWCNT, {\em i.e.}, a Josephson field
effect transistor with only two modes
\cite{Jarillosupercurrent,hij}. Experimental access to the less
transparent regimes \cite{Ralph,MSS2006Proc,vanDam,Cleuziou} is even
more interesting due to the possibility of probing the competition
between effects related to Coulomb blockade and superconductivity.
Coulomb blockade generally suppresses the supercurrent and
interesting phenomena such as $\pi$-junction behavior has been
observed in the closed quantum dot regime in nanowires
\cite{vanDam}. For more transparent devices the supercurrent can be
enhanced due to Kondo physics \cite{Glazman1989JETPL}. Both
supercurrent and Kondo physics are two extensively studied manybody
effects in condensed matter physics and SWCNT Josephson junctions
thus give a unique possibility to examine their interplay. The
supercurrent is predicted to coexist with Kondo resonances provided
that the Kondo related energy scale $k_B T_K$ is bigger than the
superconducting energy gap $\Delta$
\cite{Glazman1989JETPL,Siano2004PhRvL,Choi2004PhRvB}. The importance
of the ratio between these parameters has recently been addressed in
different measurements \cite{BuitelaarKondo}. In this article we
extend this investigation to SWCNTs contacted to superconducting
leads in order to experimentally probe the interplay between
supercurrent and Kondo physics. We investigate the gate dependence
of a narrow zero bias conductance peak interpreted as a proximity
induced supercurrent and show that the observation of a supercurrent
in Kondo resonances depends on the ratio between the two energy
scales ($k_BT_K/\Delta$) with a crossover close to 1 qualitatively
consistent with theory.

\section{Experimental Details}
SWCNTs are grown from catalyst islands consisting of Fe-oxide and
Mo-oxide supported by aluminum nano-particles \cite{KongCVD}. Growth
is performed by chemical vapor deposition at $850$\,$^\circ$C with a
controlled flow of gasses Ar: $1$\,L/min, H$_2$: $0.1$\,L/min,
CH$_4$: 0.5\,L/min. During heating the furnace is kept under an Ar
and H$_2$ flow, whereas cooling is done in Ar. We reduce cooling
time by air-cooling the furnace. The substrate is a doped silicon
wafer (used as back gate) with a $500$\,nm SiO$_2$ layer on top.
Pairs of superconducting electrodes of Ti/Al/Ti (5/40/5\,nm) are
defined directly on top of the SWCNT by electron beam lithography
followed by optical lithography to define the Cr/Au bonding pads.
The first titanium layer of the metallic trilayer ensures good
contact to the SWCNT, whereas the thicker middle aluminum layer is
the actual superconductor in the device. Finally, the top layer of
Ti is intended to stop oxidation of the aluminum. The gap between
the source and drain electrode is typically around $0.5$\,$\mu$m. In
the same evaporation process a four-probe device is made next to the
S-SWCNT-S devices, which is used to measure the transition
temperature $T_c=760$\,mK and the critical field around
$B_c=100$\,mT of the superconductor. From Bardeen-Cooper-Schrieffer
(BCS) theory we deduce the superconducting energy gap
$\Delta=1.75k_B T_c=115$\,$\mu$eV. The devices are cooled in a
$^3$He-$^4$He dilution fridge with a base temperature around
$30$\,mK and we use standard lock-in techniques.

\section{Sub-gap structure and supercurrent}
\begin{figure*}
 \center
    \includegraphics[width=0.8\textwidth]{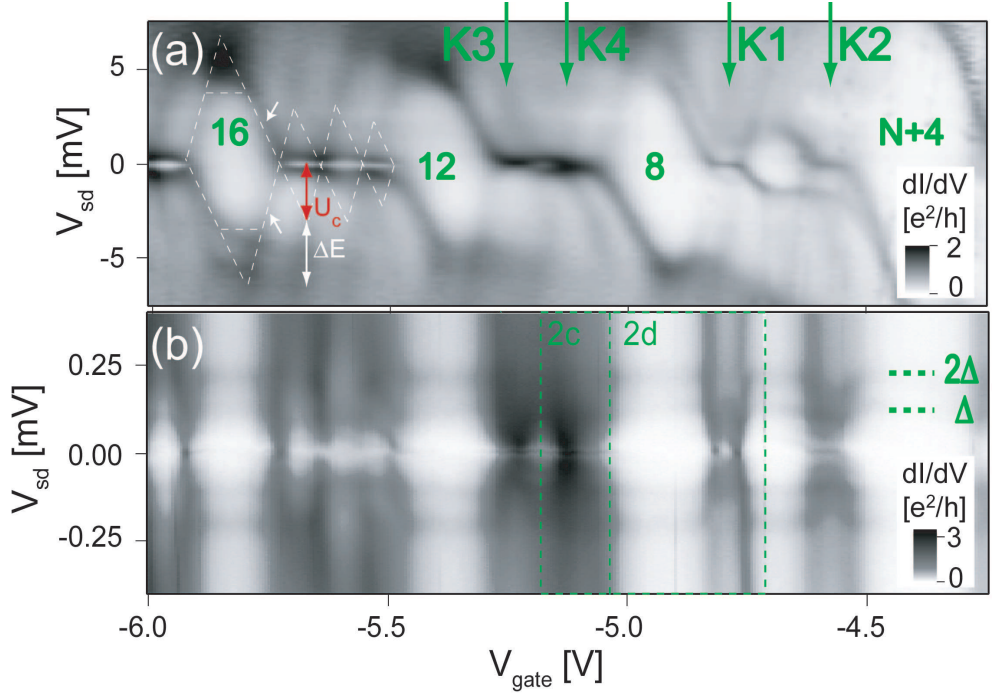}% Here is how to import EPS art
    \caption{\label{fig1} (a) Bias spectroscopy plot of
    a SWCNT device at 30\,mK showing a four-fold Coulomb blockade shell
    structure. The numbers indicate the additional hole filling, where
    big diamonds correspond to filled shells. Four Kondo resonances
    K1-K4 are identified. A magnetic field of $180$\,mT is applied to
    suppress superconductivity in the leads. (b) Bias spectroscopy plot
    for the same gate range as (a) but with the leads in the
    superconducting state. A sub-gap structure emerges, most clearly
    visible at $\pm\Delta/e$ and $\pm 2\Delta/e$ (horizontal green dashed lines). The green dashed rectangles are the gate voltage
    regions shown in Fig.\ \ref{fig2star}(c) and \ref{fig2star}(d).}
    \end{figure*}
At room temperature the current through the device is gate
dependent, which reveals that the SWCNT is semiconducting.
 Figure \ref{fig1}(a) shows a bias spectroscopy plot at 30\,mK of the SWCNT device in the Kondo regime. It
is measured at negative gate voltages and thus transport takes place
through the valence band, i.e., tunneling of holes. The
superconductivity in the leads is suppressed by a relative weak
magnetic field of $180$\,mT ($>$$B_c$). The plot shows Coulomb
blockade diamonds \cite{KouwenhovenQDReview} with a characteristic
pattern of three small diamonds followed by a bigger diamond as
indicated with white lines for hole filling 13-16. Such Coulomb
blockade diamond structure indicates a four-fold degenerate shell
structure due to spin and orbital degrees of freedom, where each
shell contains two spin-degenerate orbitals \cite{Sapmaz}. The
filled shells corresponding to the big Coulomb blockade diamonds are
marked in Fig.\ \ref{fig1}(a) by the additional number of holes on
the SWCNT quantum dot. Due to intermediate transparency contacts
significant cotunneling is allowed which tends to smear the
features. The charging energy and the level spacing between the
shells are estimated as half the source-drain height of the very
faintly visible small diamonds and by the additional source-drain
height of the big diamonds \cite{Franceschi} giving $U_c\sim 3$\,meV
and $\Delta E\sim 4$\,meV, respectively (see arrows Fig.\
\ref{fig1}). A very rough estimate of the level broadening
$\Gamma\sim 1$\,meV and the asymmetry in the coupling to source and
drain $\alpha=\Gamma_s/\Gamma_d\sim 0.4$ are extracted from the
current plateaus at negative and positive bias (100\,nA and -60\,nA)
at the white arrows. The current is modified due to the four-fold
degeneracy which enhances the in-tunneling rate with 4 compared to
the out-tunneling rate giving
$(e/\hbar)4\Gamma_s\Gamma_d/(4\Gamma_s+\Gamma_d)$ and
$(e/\hbar)(4\Gamma_d \Gamma_s/(\Gamma_s+4\Gamma_d)$ for negative and
positive bias polarity, respectively \cite{Babic}. Furthermore, four
Kondo resonances K1-K4 are identified (green arrows) and labeled in
ascending order based on their Kondo temperatures. The Kondo
temperatures are estimated by fitting the normal state conductance
versus $V_{sd}$ (solid line in Fig.\ \ref{fig2}a-b) to a Lorentzian
line shape. The half width at half maximum of each Lorentzian fit
yields the estimated Kondo temperature giving $T_K\sim 2$\,K,
4.5\,K, 5\,K and 6\,K for K1-K4. For hole filling 13-15, conductance
ridges at low finite bias are seen instead of zero bias Kondo
resonances. The exact origin of these features are not fully
understood, but they are attributed to inelastic cotunneling through
two slightly split orbitals in the shell, which might include Kondo
physics as well \cite{Paaske}. We note that these lines are not
related to superconductivity.

Figure \ref{fig1}(b) shows a bias spectroscopy plot at low bias
voltages for the same gate range as Fig.\ \ref{fig1}(a) with the
leads in the superconducting state. A sub-gap structure clearly
appears \cite{Buitelaar}. The peaks in differential conductance at
$V_{sd}=\pm 2\Delta/e\sim \pm 230$\,$\mu$V are attributed to the
onset of quasi-particle tunneling. At lower bias the transport is
governed by Andreev reflections, which are possible due to the
intermediate transparency between the SWCNT and the superconducting
leads \cite{Andreev}. Features at biases $V_{n} =\pm\ 2\Delta/(e
n),n=2,3,...$ are expected due to the opening of higher order
multiple Andreev reflection processes as the bias is lowered
\cite{klapwijk,bratus}. Peaks at $V_{sd}=\pm \Delta/e\sim \pm
115$\,$\mu$V are clearly seen consistent with the energy gap found
from BCS theory. Furthermore, for some ranges in gate voltage a
narrow zero bias conductance peak is seen. We note, that this zero
bias peak is visible in most of the Kondo resonances and is the
subject to analysis below.

\begin{figure}
\center
    \includegraphics[width=0.75\textwidth]{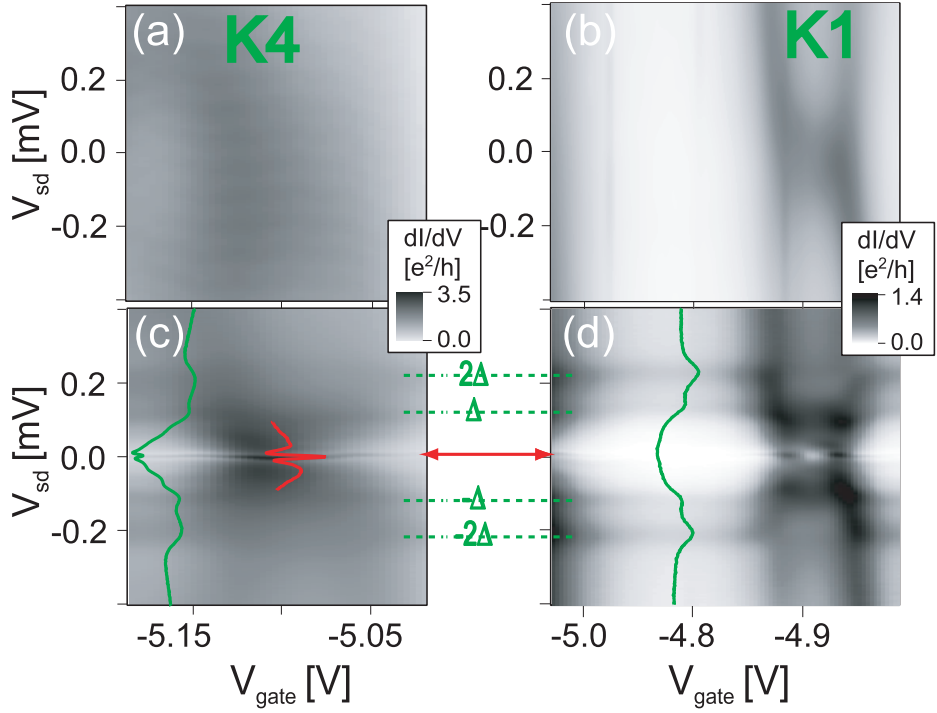}% Here is how to import EPS art
    \caption{\label{fig2star} (a-b) Bias spectroscopy plot at 30\,mK with the leads in the normal state
    ($B=180$\,mT) for Kondo resonance K4 and K1, respectively. (c-d) High resolution data with the leads superconducting for the
same gate and bias voltages as in the normal state above. The
corresponding gate regions are shown by the dashed green rectangles
in Fig.\ \ref{fig1}(b). A sub-gap structure appears at bias voltage
$\pm2\Delta/e$ and $\pm\Delta/e$ indicated by the green dashed
lines. This is particular clear in the Coulomb blockade region (d)
and also seen in the green traces, which are $dI/dV$ versus bias
curves extracted from the plots. A zero bias conductance peak (red
arrow) is also visible throughout Kondo resonance K4 as shown by the
red trace in (d) and for some gate voltages in (c).}
\end{figure}
These above mentioned effects due to superconductivity are more
clearly revealed in high resolution data shown in Fig.\
\ref{fig2star}(c) and \ref{fig2star}(d) corresponding to the green
dashed rectangles in Fig.\ \ref{fig1}(b). Figure \ref{fig2star}(a)
and \ref{fig2star}(b) show the same gate and bias regions as (c) and
(d) with the leads in the normal state for comparison. Clearly the
sub-gap structure is due to the proximity of the superconducting
leads. The peaks at $\pm2\Delta/e$ and $\pm \Delta/e$ are mostly
pronounced in the Coulomb blockade region as expected due to lower
effective transparency \cite{klapwijk} (d) while being more smeared
in the high conducting region of K4 (c). Green curves show bias
cuts, where higher order multiple Andreev reflection features are
faintly visible in (c). A strong gate dependence of the sub-gap
structure is observed when the gate voltage is tuned closer to the
Coulomb resonances on each side of K1 from the Coulomb diamond with
even occupation. This behavior has previously been observed
\cite{Buitelaar} and explained by the interplay between multiple
Andreev reflection and a resonant level
\cite{Yeyati1997PhRvB,Johansson1999PhRvB}. The zero bias conductance
peak indicated by the red arrow is clearly visible in Kondo
resonance K4 (red trace in (c)) and also present for some gate
voltages in (d). Similar behavior of a zero bias peak in Kondo
resonances have been observed in other devices \cite{MSS2006Proc}.
We interpret the zero bias conductance peak as being due to a
proximity induced supercurrent running through the SWCNT. The
magnitude of the supercurrent is estimated by the area of the peak
giving a typical value in the order of $I_m\sim 0.2$\,nA. A similar
analysis of a zero bias conductance peak as supercurrent has
successfully been carried out for an open quantum dot \cite{hij},
{\em i.e.}, without Coulomb blockade effects. We note that an
alternative interpretation of the origin of the zero bias peak based
on quasiparticle current (multiple Andreev reflections) with a
cut-off for higher order multiple Andreev reflection processes has
also been suggested \cite{vecino}, but in this article the
supercurrent interpretation will be pursued. The value of the
measured supercurrent is highly suppressed compared to its
theoretically expected value for one spin-degenerate level with the
deduced asymmetric coupling ($\alpha=0.4$) at resonance
$I_0^{res}=2e\Delta/\hbar~(\Gamma_{min}/\Gamma) \sim 8$\,nA in the
broad resonance (short) regime $\Gamma>\Delta$, where
$\Gamma_{min}=min(\Gamma_s,\Gamma_d)\sim 0.3$\,meV
\cite{Beenakker2004condmat}. Despite having two spin-degenerate
levels in the SWCNT, only one level is available due to Coulomb
blockade.

The suppression is partly explained within an extended resistively
and capacitively shunted junction model due to interaction of the
SWCNT Josephson junction with its electrical environment. The
electrical environment for this device is the low (serial)
resistance between the Josephson junction and the bonding pads as
well as the capacitance between the source and drain bonding pads
via the backgate. The low serial resistance turns out to play a
crucial role making the device an {\it underdamped} Josephson
junction \cite{hij,Steinbach,Vion,Jarillosupercurrent}. This means
that the full value of the critical current is never measured due to
thermally activated phase-slips. A more intuitive understanding is
obtained by the mechanical analog to the Josephson junction of a
fictious particle in a tilted washboard potential, where the tilt is
given by the current and the phase difference between the
superconductors is "running" as the fictious particle moves in the
potential. When the particle stays in a minimum of the washboard
potential (constant phase), a supercurrent is seen. However, as the
current increases the washboard potential is tilted and at the
critical current the particle can slide into the next minimum. If
thermal excitations are significant this process happens at lower
current than the critical current. Since the junction is
underdamped, the friction of the particle is low and it thus easily
acquires a "run away" phase, that suppresses the supercurrent, which
results in a much lower measured supercurrent. In contrast to the
reported underdamped carbon nanotube Josephson junction in the
Fabry-Perot regime \cite{hij}, the effect of Coulomb blockade and
single hole tunneling also contribute to the suppression of the
supercurrent, because Cooper pair transport is a two particle
tunneling process. A quantitative analysis of the suppression due to
Coulomb blockade is outside the scope of this article. This issue is
being addressed in a separate work, where the on-chip electrical
circuit of the SWCNT Josephson junction has been modified and the
effects of Coulomb blockade thus can be compared to theoretical
calculations \cite{pi}. Finally, we note that noise effects giving
rise to thermally activated phase slips can be analyzed in the
overdamped case \cite{Ambegaokar}.

\section{Kondo physics and supercurrent}
\begin{figure}[t!]
{\center
    \includegraphics[width=0.8\textwidth]{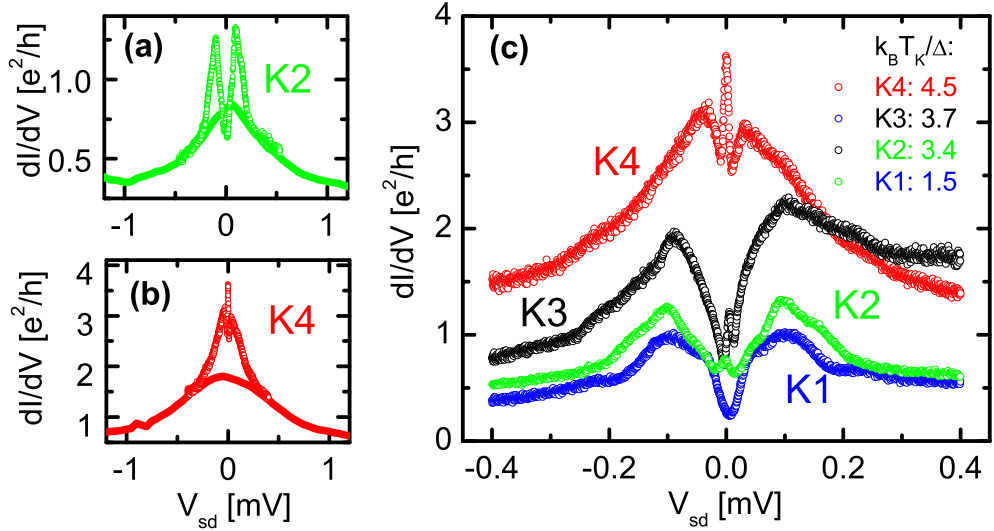}% Here is how to import EPS art
    \caption{\label{fig2} $dI/dV$ versus $V_{sd}$ for Kondo resonances K1-K4 shown in Fig.\
    \ref{fig1}(a)
    illustrating the effect of superconducting leads. (a-b) Solid lines and circles are with the
    leads in the normal and superconducting state, respectively. (c) The four Kondo resonances K1-K4
    for small bias voltages with the leads in the superconducting state. The measured supercurrent given by the zero bias peak area
    is zero for the lowest value of the ratio $k_BT_K/\Delta$, while it increases as this ratio is increased.
    }}
\end{figure}

We now return to the Kondo resonances K1-K4 and analyze the
interplay between Kondo physics and supercurrent. The solid curves
in Fig.\ \ref{fig2}(a) and \ref{fig2}(b) show bias cuts with the
leads in the normal state through the center of Kondo resonance K2
and K4, respectively (see Fig.\ \ref{fig1}(a)). The circles
correspond to the behavior when the leads are superconducting. In
both cases an enhancement of the differential conductance is
observed for bias voltages between $\pm 2\Delta/e$ and a zero bias
conductance peak is present due to supercurrent. The measured
supercurrent is largest for the Kondo resonance with the highest
Kondo temperature (K4), {\em i.e.}, broadest Kondo resonance. In
Fig.\ \ref{fig2}(c) bias sweeps at the center of all four Kondo
resonances are shown with the leads in the superconducting state.
The Kondo temperature normalized by the superconducting energy gap
($k_B T_K/\Delta$) is the important parameter and is given for each
resonance in the figure. It is seen that the measured supercurrent
vanishes for the lowest ratio (blue circles) while it emerges and
increases as the ratio is increased. The crossover is close to
$k_BT_K/\Delta\sim 1$.

\begin{figure*}[t!]
    \center
    \includegraphics[width=0.98\textwidth]{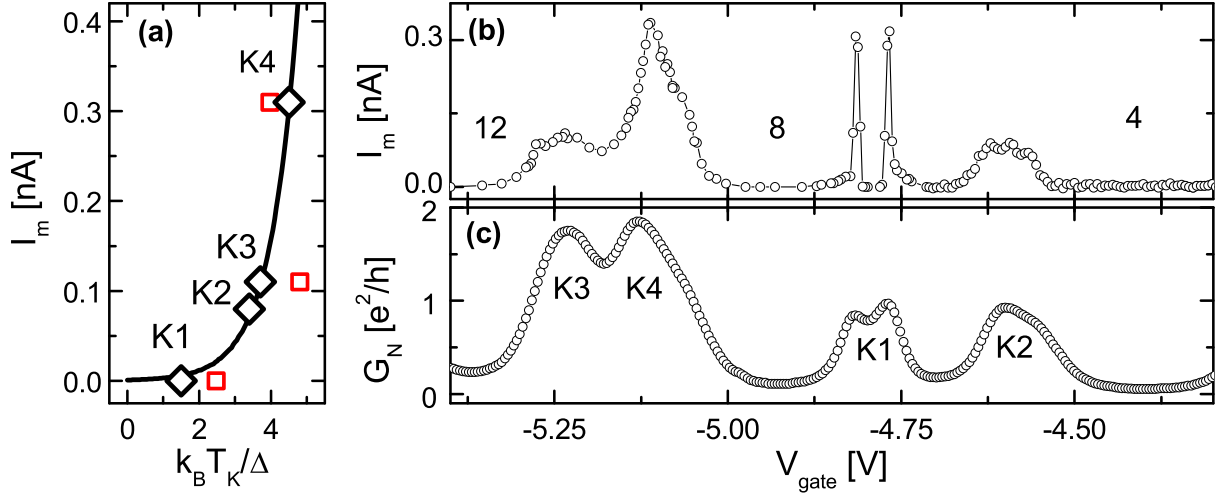}% withonlyres Here is how to import EPS art
    \caption{\label{fig:PeakvsVgate} (a) The measured supercurrent is zero for K1 and shown to increase as a function of
    $k_BT_K/\Delta$ (K1-K4) . An exponential fit to the data points is given
    by the solid curve as guideline, where the Kondo temperature is based on the width of the resonance (black diamonds).
    The red squares are the Kondo temperature obtained from the temperature dependence
    (K1, K3 and K4). (b) Measured supercurrent versus gate voltage in the range with the Kondo resonances K1-K4,
    where the additional hole number is given for filled shells as in Fig.\ \ref{fig1}(a).
     A finite supercurrent is seen in the center of the broader Kondo resonance K2, while being zero in the center of narrowest Kondo resonance
     K1. This behavior does not reflect the normal state conductance
     $G_N$ shown for the same gate range below in (c).}
    \end{figure*}
To illustrate this point more clearly, the measured supercurrent
versus $k_BT_K/\Delta$ is plotted in Fig.\ \ref{fig:PeakvsVgate}(a)
for the four Kondo resonances analyzed above (black diamonds). The
solid curve shows a guideline to the eye based on an exponential fit
to the available data points. The overall trend is qualitative
consistent with existing theory, which predicts a suppression of the
supercurrent in the socalled weak coupling regime ($k_BT_K \ll
\Delta$), while the supercurrent coexists with Kondo resonances in
the strong coupling regime ($k_BT_K\gg\Delta$)
\cite{Choi2004PhRvB,Siano2004PhRvL}. The red squares show the
measured supercurrent where the Kondo temperature is extracted from
the temperature dependence for completeness (only data available for
K1, K3 and K4) \cite{GoldhaberGordon1998Nature}.

Finally, Fig.\ \ref{fig:PeakvsVgate}(b) shows the measured
supercurrent as a function of gate voltage in the range including
the four Kondo resonances. It is strongly gate dependent
illustrating effects of the four Kondo resonances. Figure
\ref{fig:PeakvsVgate}(c) shows the zero bias conductance $G_N$ with
the leads in the normal state for the same gate range. The
supercurrent {\em does not} directly reflect the normal state
conductance $G_N$ as in the case of an open quantum dot
\cite{Jarillosupercurrent,hij}, where the supercurrent is uniquely
determined by the normal state conductance. This is most clearly
seen by comparing Kondo resonance K1 and K2. No supercurrent is
present in the center of the Kondo resonance K1 despite the high
normal state conductance, while a finite supercurrent is present in
Kondo resonance K2 with equally high normal state conductance in
contrast to the behavior expected in the open quantum dot regime.
Similarly, the Kondo resonances K3 and K4 have almost the same
normal state conductances, but very different measured supercurrent.
These observations support the above analysis that the interplay
between Kondo physics and superconductivity has the fraction
$k_BT_K/\Delta$ as the important parameter for the observation of
supercurrent in Kondo resonances, i.e., the bias width of the Kondo
resonances and not only their normal state conductance is important.
We also note that the measured supercurrent in diamond 4, 8 and 12
is zero due to Coulomb blockade. The supercurrent in the high $T_K$
Kondo resonances can thus be view as being enhanced from the
suppressed values by Coulomb blockade and the single spin due to the
formation of Kondo resonances. Furthermore, the overall behavior of
the supercurrent of these four Kondo resonances also indirectly
indicates a $\pi$ to 0 transition of the current phase relation of
the Josephson junction as a function of the parameter $k_B
T_K/\Delta$ [17,18].  Similar gate voltage behavior has been
observed in Ref.\ \cite{Cleuziou}, but the authors do not show the
magnitude of the supercurrent as a function of Kondo temperature. We
end by noting that for large Kondo temperatures compared to the
superconducting energy gap (K4), the observed gate voltage
dependence of the supercurrent resembles the behavior of a
superconducting single electron transistor, i.e., the supercurrent
obtains a sharp maximum at the gate voltage corresponding to the
center of the odd diamond \cite{Joyez}.

\section{Conclusion}
In conclusion SWCNTs have been contacted to superconducting Ti/Al/Ti
leads creating SWCNT Josephson junctions. We observe sub-gap
structure due to multiple Andreev reflections and a narrow zero bias
conductance peak. The zero bias peak is interpreted as a proximity
induced supercurrent with a suppressed magnitude due to the
underdamped nature of the Josephson junction and Coulomb blockade.
We examine its interplay with Kondo resonances and the measured
supercurrent is shown to coexist with Kondo resonances which have
high Kondo temperatures compared to the superconducting energy gap,
while being suppressed when the Kondo temperatures becomes
comparable with the superconducting energy gap, qualitative
consistent with existing theory.

\section*{Acknowledgement}
We like to thank Karsten Flensberg, Tom{\'a}\v{s} Novotn{\'y} and
Jens Paaske for fruitful discussions. Furthermore, we wish to
acknowledge the support of the Danish Technical Research Council
(The Nanomagnetism framework program), EU-STREP Ultra-1D and CARDEQ
programs.

\section*{References}
\bibliography{bibl}% Produces the bibliography via BibTeX.
\bibliographystyle{unsrt}

\end{document}